\documentclass[aps,twocolumn,preprintnumbers,nofootinbib,superscriptaddress]{revtex4}
\usepackage{amsmath} \usepackage{graphicx} \usepackage{amsfonts}
\usepackage{array} \usepackage{amsthm} \usepackage{bm}
\usepackage{palatino} \usepackage{mathpazo} 
\usepackage{supertabular}

\usepackage[breaklinks]{hyperref}
\usepackage{color}

\newcommand{\nn}{\nonumber}
\newcommand{\be}{\begin{equation}}
\newcommand{\ee}{\end{equation}}
\newcommand{\ba}{\begin{eqnarray}}
\newcommand{\ea}{\end{eqnarray}}
\newcommand{\bal}{\begin{align}}
\newcommand{\eal}{\end{align}}

\newcommand{\e}{{\rm e}}
\newcommand{\dd}{{\rm d}}

\newcommand{\bb}{\bibitem}

\newcommand{\om}{\omega}
\newcommand{\al}{\alpha}

\newcommand{\bt}{\beta}

\newcommand{\ga}{\gamma}
\newcommand{\ro}{\rho}
\newcommand{\ep}{\epsilon}

\newcommand{\ta}{\theta}

\newcommand{\Si}{\Sigma}
\newcommand{\De}{\Delta}

\newcommand{\Om}{\Omega}

\newcommand{\de}{\delta}

\newcommand{\bw}{\begin{widetext}}
\newcommand{\ew}{\end{widetext}}

\begin{document}
\title{Wormhole solutions sourced by fluids, II: Three-fluid two-charged sources}

\author{Mustapha Azreg-A\"{\i}nou}
\affiliation{Ba\c{s}kent University, Faculty of Engineering, Ba\u{g}l\i ca Campus, 06810 Ankara, Turkey}


\begin{abstract}
Lack of a consistent metric for generating rotating wormholes motivates us to present a new one endowed with interesting physical and geometrical properties. When combined with the generalized method of superposition of fields, which consists in attaching a form of matter to each moving frame, it generates massive and charged (charge without charge) two-fluid-sourced, massive and two-charged three-fluid-sourced, rotating as well as new static wormholes which, otherwise, can hardly be derived by integration. If the lapse function of the static wormhole is bounded from above, no closed timelike curves occur in the rotating counterpart. For positive energy densities dying out faster than $1/r$, the angular velocity includes in its expansion a correction term, to the leading one that corresponds to ordinary stars, proportional to $\ln r/r^4$. Such a term is not present in the corresponding expansion for the Kerr-Newman black hole. Based on this observation and our previous work, the dragging effects of falling neutral objects may constitute a substitute for other known techniques used for testing the nature of the rotating black hole candidates that are harbored in the center of galaxies. We discuss the possibility of generating ($n+1$)-fluid-sourced, $n$-charged, rotating as well as static wormholes.
\end{abstract}


\maketitle

\section{Introduction\label{secI}}

In order to elucidate the nature of the black hole candidates at the center of galaxies workers use different theoretical approaches and techniques~\cite{Sgr}-\cite{types} among which we find imaging, that is, the observation of the shadow of the hole in the sky. We have shown that imaging, applied to nonrotating solutions, remains inconclusive to whether the black hole candidate, located at Sagittarius A$^\star$ (Sgr A$^\star$), is a supermassive black hole or a supermassive type I wormhole~\cite{types}. Recall that type I wormholes are the solution that violate the least the local energy conditions.

Some other tests apply exclusively to rotating solutions~\cite{X-ray}. The only rotating black hole solutions to the field equations of general relativity are the Kerr and the Kerr-Newman black holes. Rotating wormhole solutions to the same field equations exist, among which the one derived in~\cite{teo}, which has been shown to be sourced by two unphysical fluids~\cite{Trot}, and the charged solutions derived in~\cite{Trot}. All solutions derived in Refs.~\cite{teo,Trot} suffer from the strong assumption neglecting flattening due to rotation and they may remain valid only in the slow rotation limit.

Among tests that apply exclusively to rotating solutions is the dragging of neutral objects. The dragging effects of the rotating wormhole derived in Ref.~\cite{teo} are, by construction, those of ordinary stars. The angular velocity of the massive and charged rotating wormhole derived in Ref.~\cite{Trot} has a series expansion depending on both the mass $M$ and charge $Q$ but, unlike the Kerr-Newman black hole, it does not include the term $1/r^4$ for all values of ($M,Q$). Moreover, there exists a mass-charge constraint yielding almost no more dragging effects than ordinary stars. From these results, we see how the dragging effect can be used as a substitute test for elucidating the nature of the black hole candidates at the center of galaxies. This conclusion will be confirmed in this work.

In this work we intend to drop the non-flattening constraint in the hope to obtain more realistic solutions. We introduce a simple definition of the flattening condition and observe it throughout the paper. Since there is a growing interest in obtaining \emph{analytical} rotating wormhole solutions for their use in astrophysics, we focus in this work on rotating and non-rotating (static) wormholes. Based on our previous work~\cite{az3,AzregR}, we derive from our formula developed therein, which is intended to generate all types of rotating solutions, a Kerr-like metric for generating rotating wormholes from their known non-rotating counterparts. As a bonus, the formula works the other way around and it allows to construct new static wormholes as being the limit $a\to 0$ of their rotating counterparts, where $a$ is the rotation parameter.

A couple of other exact, numerical, as well as slowly, rotating wormhole solutions were also derived~\cite{khat}-\cite{refw}. In this work, we rather derive families of one-, two-, and three-fluid-sourced rotating and new static wormholes. In order to achieve that in a systematic approach, in Secs.~\ref{secnrw} and~\ref{secrw} we review the geometries of nonrotating wormholes and rotating stars where we focus more on rotating wormholes. The generic rotating metric depending on two unknown functions ($A,b$) and on $a$ is derived in Sec.~\ref{secklm}. Here ($A,b$) is the metric of the static wormhole in Schwarzschild coordinates in the notation of Morris and Thorne~\cite{MT}. The geometrical and physical properties of the rotating wormhole metric along with a definition of the flattening constraint are discussed in Sec.~\ref{geo-phy}.

Sections.~\ref{secof}, \ref{sectf}, and~\ref{sec3f} are respectively devoted to the derivations and analytical discussions of the properties of the families of massive one-fluid-sourced, massive and charged two-fluid-sourced, and massive and two-charged three-fluid-sourced rotating, and their static counterpart, wormholes.  The charge refereed to in this work, being either electric or magnetic, is attached to a source-free electromagnetic field. This is the well-known Misner-Wheeler effect ``charge without charge"~\cite{MW}. In Sec.~\ref{secwec} we address the question of the local energy conditions. In Sec.~\ref{secnf} we generalize the approach of the superposition of fields to lead to ($n+1$)-fluid-sourced, $n$-charged (massless and massive), rotating and static wormholes and we conclude in Sec.~\ref{secc}.

\section{Geometry of the spacetime of a nonrotating wormhole \label{secnrw}}

The geometry of the spacetime of a nonrotating wormhole is better described by the Morris and Thorne metric~\cite{MT}
\begin{equation}\label{g1}
    \dd s^2=A(r)\dd t^2-\frac{\dd r^2}{1-b(r)/r}-r^2\dd \Omega^2,
\end{equation}
in Schwarzschild coordinates where $A$ is the lapse function and $b$ is the shape one. The throat is the sphere of equation $r=r_0=b(r_0)$. For simplicity, we assume symmetry of the two asymptotically flat regions, which particularly implies that if the mass of the wormhole is finite then it is the same as seen from both spatial infinities. The functions $A$ and $b$ are constrained by~\cite{MT,V}
\begin{align}
&\lim_{r\to\infty}A=\text{finite}=1,\nn\\
&b<r\text{ if }r>r_0\;\text{ and }\;b(r_0)=r_0,\nn\\
\label{g2}&\lim_{r\to\infty}(b/r)=0,\\
&rb'<b\;(\text{near the throat}),\nn\\
&b'(r_0)\leq 1.\nn
\end{align}
In this paper a prime notation $f'(r,\ta,\cdots)$ denotes the partial derivative of $f$ with respect to (wrt) $r$, and derivation wrt to other variables is shown using the indexical notation, as in $f_{,\ta}\equiv \partial f/\partial \ta$. The value of the limit in the first line~\eqref{g2} is set to 1 by rescaling $A$ and redefining $t$. If the mass of the wormhole is finite, we have the further constraint
\begin{equation}\label{g3}
\lim_{r\to\infty}b\equiv b_{\infty}=2GM=2M.
\end{equation}

The SET is usually taken anisotropic of the form~\cite{MT} $T^{\mu}{}_{\nu}={\rm diag}(\ep(r),-p_r(r),-p_t(r),-p_t(r))$, $\ep$ being the energy density and $p_r$ and $p_t$ are the radial and transverse pressures. The filed equations $G^{t}{}_{t}=8\pi T^{t}{}_{t}$, $G^{r}{}_{r}=8\pi T^{r}{}_{r}$, and the identity $T^{\mu}{}_{r;\mu}\equiv 0$ yield, respectively
\begin{align}\label{g3b}
&b'=8\pi r^2\ep ,\nn\\
&(\ln A)'=\frac{8 \pi  r^3p_r +b}{r (r-b)},\\
&4 p_t=4 p_r+2 r p_r{}'+r(p_r+\ep )(\ln A)'.\nn
\end{align}

\section{Rotating geometries\label{secrw}}

\subsection{The standard metric of a rotating star}

The standard metric of a circular, stationary, and axisymmetric spacetime, admitting two commuting Killing vectors $\partial_t$ and $\partial_{\phi}$, may be brought to the following form in quasi-isotropic coordinates~\cite{rot1,rot2} (see~\cite{Eric} for more details):
\begin{equation}\label{g4}
    \dd s^2=N^2\dd t^2-D_1^2(\dd R^2+R^2\dd\ta^2)-D_2^2R^2\sin^2\ta(\dd\phi-\om\dd t)^2.
\end{equation}
Note that there is no restriction in having
\begin{equation}\label{g5}
g_{\ta\ta}/g_{RR}=R^2,
\end{equation}
as in~\eqref{g4}; rather, this reflects the fact that all two-dimensional metrics are related by a conformal factor. Here ($N^2,D_1^2,D_2^2,\om$) are positive functions depending on ($R,\ta$).

The form~\eqref{g4} is not convenient for constructing wormhole or black hole solutions~\cite{Trot}. Introducing a new radial coordinate $r$
\begin{equation}\label{g6}
    R\equiv R(r),
\end{equation}
we bring it to the form
\begin{equation}\label{g7}
    \dd s^2=N^2\dd t^2-\e^{\mu}\dd r^2-r^2K^2[\dd\ta^2+F^2\sin^2\ta(\dd\phi-\om\dd t)^2],
\end{equation}
where
\begin{equation*}
\e^{\mu(r,\ta)}=\Big(D_1\frac{\dd R}{\dd r}\Big)^2,\; r^2K^2(r,\ta)=D_1^2R^2,\; F^2(r,\ta)=\frac{D_2^2}{D_1^2}.
\end{equation*}
The property~\eqref{g5} is lost in~\eqref{g7} but the ratio
\begin{equation}\label{g7a}
g_{\ta\ta}/g_{rr}=(\dd\ln R/\dd r)^{-2}
\end{equation}
is still independent of $\ta$.

If the star rotates slowly its shape is not flattened by the centrifugal forces, so it retains its spherical symmetry resulting in $g_{\phi\phi}=g_{\ta\ta}\sin^2\ta$\cite{Eric}, that is, in $F^2=1$. Rotating wormholes satisfying the no-flattening condition $F^2\equiv 1$ were derived in~\cite{teo,Trot}. For fluid-sourced stars where the angular velocity is differential vanishing at spatial infinity as the inverse cubic power of the radial distance, a natural flattening condition would be
\begin{equation}\label{g8}
    F^2\geq 1,
\end{equation}
where the saturation is attained on the axis of rotation ($\ta=0$ or $\ta=\pi$) and at spatial infinity where the centrifugal forces tend to vanish.

\subsection{The standard metric of a rotating wormhole}

If intended for the derivation of rotating wormholes, the metric coefficient $-g_{rr}$~\eqref{g7} is preferably brought to the Morris and Thorne form
\begin{equation}\label{g9}
\e^{\mu(r,\ta)}\equiv \frac{1}{1-B(r,\ta)/r},
\end{equation}
where we use $B(r,\ta)$ for rotating wormholes and $b(r)$ for nonrotating ones. The surface of the throat is defined by
\begin{equation}\label{g10}
B(r_0,\ta_0)=r_0.
\end{equation}
This, in general, provides $r_0$ as a function of $\ta_0$; that is, for a given value of $\ta_0$ we solve~\eqref{g10} for $r_0$ and we keep its largest value.

The metric~\eqref{g7} is not free of singularities unless it fulfils some conditions. In the no-flattening case $F^2\equiv 1$, the curvature and Kretschmann scalars contain the expression $r-B$ in their denominators. Assuming $B'|_{(r_0,\ta_0)}\neq 1$, it was shown that the first three partial derivatives of $B$ wrt to $\ta$ must vanish on the throat~\cite{teo,Lobo,Trot}
\begin{equation}\label{g11}
B_{,\ta}|_{(r_0,\ta_0)}=0,\; B_{,\ta\ta}|_{(r_0,\ta_0)}=0,\; B_{,\ta\ta\ta}|_{(r_0,\ta_0)}=0,
\end{equation}
for the two scalar invariants to have well defined values on the throat and off it~\cite{Trot}. This conclusion extends to the flattening case~\eqref{g8}.

If $F^2\neq 1$, the curvature scalar $\mathcal{R}$ has, besides the denominator $r-B$, the denominator
\begin{equation}\label{nd}
    F^2.
\end{equation}
In a more general physical configuration, not obeying~\eqref{g8}, where $F^2$ is allowed to vanish on the throat or off it, other constraints than~\eqref{g11} must be imposed to ensure regularity of the two scalar invariants. We will not pursue this discussion here for it does not concern us for the remaining parts of this work.

\subsection{A Kerr-like metric for rotating wormholes\label{secklm}}

We intend to use a metric that guarantees in its generality the regularity of the above-mentioned scalar invariants with no constraints and retains the flattening condition~\eqref{g8}. The metric has been used in Ref.~\cite{AzregR} to generate rotating regular black holes, in Ref.~\cite{az3} to generate fluid wormholes with and without electric or magnetic field, and in Ref.~\cite{conf} to generate regular cores. It has the following Kerr-like form
\begin{align}
&\dd s^2 =\Big(1-\frac{2f}{\ro^2}\Big)\dd t^2-\frac{\ro^2 A}{\De}\,\frac{\dd r^2}{1-b/r}\nn\\
\label{rm}&+\frac{4af \sin ^2\theta}{\ro^2}\,\dd t\dd \phi-\ro^2\dd \ta^2-\frac{\Si\sin ^2\theta}{\ro^2}\,\dd \phi^2,\\
&\ro^2\equiv r^2+a^2 \cos^2\ta,\quad 2f(r)\equiv r^2(1-A)\nn\\
&\De(r)\equiv r^2 A+a^2,\quad \Si\equiv (r^2+a^2)^2-a^2\De\sin^2\ta \nn,
\end{align}
which reduces to~\eqref{g1} in the limit of no rotation. The metric~\eqref{rm} is derived upon first transforming the static metric~\eqref{g1} into a form where $g_{tt}=1/g_{\bar{r}\bar{r}}$. This is achieved using the new radial coordinate $\bar{r}$ defined by $\dd \bar{r}^2=A\dd r^2/(1-b/r)$. Upon omitting the factor\footnote{The factor $\Psi/\ro^2$, used in Refs.~\cite{AzregR,az3,conf}, is to ensure that the rotating metric is sourced by \emph{one} fluid. In this work we will set other conditions, constraining $A$ and $b$, to derive one-, two-, and three-fluid sourced rotating solutions.} $\Psi/\ro^2$, Eqs.~(16) and~(18) of Ref.~\cite{AzregR} yield the same metric~\eqref{rm} with the second term replaced by
\begin{equation*}
 -\frac{\ro^2}{\De}\,\dd \bar{r}^2,
\end{equation*}
and the function ($\ro^2,f,\De,\Si$) are expressed as in~\eqref{rm} with $r^2:=r^2(\bar{r})$. Changing back to $r$, we obtain~\eqref{rm}.

The metric~\eqref{rm} is cast into three other equivalent forms
\begin{align}
\label{rm1}\dd s^2 =&\frac{\De}{\ro^2}(\dd t-a\sin^2\ta\dd \phi)^2-\frac{\ro^2 A}{\De}\,\frac{\dd r^2}{1-b/r}-\ro^2\dd \ta^2\nn\\
&-\frac{\sin^2\ta}{\ro^2}\,[(r^2+a^2)\dd \phi-a\dd t]^2,
\end{align}
\begin{align}
\label{rm2}\dd \bar{s}^2 =&\frac{\ro^2\De}{\Si}\,\dd t^2-\frac{\ro^2 A}{\De}\,\frac{\dd r^2}{1-b/r}
-\ro^2\dd \ta^2\nn\\
&-\frac{\Si\sin^2\ta}{\ro^2}\,(\dd \phi-\om\dd t)^2\qquad \Big(\om\equiv \frac{2af}{\Si}\Big),\\
=&\frac{\ro^2\De}{\Si}\,\dd t^2-\frac{\ro^2 A}{\De}\,\frac{\dd r^2}{1-b/r}
-\ro^2\dd \ta^2\nn\\
&-\frac{\sin^2\ta}{\ro^2\Si}\,(\Si\dd \phi-2af\dd t)^2,\nn
\end{align}
\begin{align}
&\dd \tilde{s}^2 =\frac{1}{\ro^2\De_{\ta}}\,(\De_{\ta}\dd t+2af\sin^2\ta\dd \phi)^2-\frac{\ro^2 A}{\De}\,\frac{\dd r^2}{1-b/r}\nn\\
&\label{rm3}-\ro^2\dd \ta^2-\frac{\ro^2\De\sin^2\ta}{\De_{\ta}}\,\dd \phi^2,\\
&\De_{\ta}\equiv r^2A+a^2\cos^2\ta.\nn
\end{align}
The new function $\De_{\ta}$ is related to $\De$, $\ro^2$, and $f$ by: $\De_{\ta}=\De - a^2\sin^2\ta=\ro^2-2f$.

This way of casting a given rotating metric constitutes, as we shall see in the subsequent sections, our method for constructing one-, two-, and three-fluid sourced (wormhole or other) solutions.

\section{Geometrical and physical properties of the rotating wormhole \label{geo-phy}}

The best way to justify the metric~\eqref{rm} is through its physical and geometrical properties. In its full generality (no constraint however that may be on $A$ and $b$), the metric~\eqref{rm} is promising as it satisfies nice physical properties and obeys the following desired requirements.
\begin{itemize}
  \item From~\eqref{rm2} we see that the ratio $g_{\ta\ta}/g_{rr}$ is independent of $\ta$ as in~\eqref{g7a}. This implies that the metric~\eqref{rm2}, which is a special case of~\eqref{g7}, fulfils the required symmetry properties of a stationary and axisymmetric spacetime that is circular (admitting the existence of two commuting Killing vectors $\partial_t$ and $\partial_{\phi}$).

  \item Since $g_{\ta\ta}=-\ro^2$ and $g_{\phi\phi}/\sin^2\ta=-\Si/\ro^2$ are equal on the axis of rotation ($\ta=0$ or $\ta=\pi$)~\cite{Eric}, the metric~\eqref{rm} has no conical singularity on it.

  \item The metric function $-g_{rr}$, if brought to the Morris-Thorne form as in~\eqref{g9}, defines the shape function $B(r,\ta)$ of the rotating wormhole~\eqref{rm} by:
\begin{equation}\label{p1}
1-B(r,\ta)/r =   \frac{\De}{\ro^2 A}\,(1-b/r).
\end{equation}
Recall that the surface of the throat is defined by the equation $B(r_0,\ta_0)=r_0$, which in general provides $r_0$ as a function of $\ta_0$.  The lhs of~\eqref{p1} vanishes on the throat, but since $\De(r_0)\neq 0$, this implies $b(r_0)=r_0$. Thus, the throat is the \emph{nonspherical} surface of revolution whose points are located at a fixed value of the radial coordinate $r=r_0$ that is independent of the value of $\ta$. The shape of the throat is determined by the function $\ro^2$ which measures the square of the proper radial distance. On the throat, this function increases from $r_0^2$, on the equatorial plane, to $r_0^2+a^2$, on the axis of rotation.

  \item From~\eqref{p1}, it is easy to establish that the $n$th derivative of the shape function $B$ wrt $\ta$ is proportional to $1-b/r$; thus, all partial derivatives of $B$ wrt $\ta$ vanish at the throat. This guarantees that the curvature scalar $\mathcal{R}$ is finite everywhere. Direct derivations show that
\begin{equation}\label{sca}
    \mathcal{R}=P_{C}/(r\ro^6A^2),
\end{equation}
where $P_C$ is a polynomial in $A(r)$ and its first and second derivatives, $b(r)$ and its first derivative, $r^2$, $\cos^2\ta$, and $a^2$.

Equation~\eqref{sca} shows that the denominator~\eqref{nd} is not a pole of $\mathcal{R}$ if the geometry is described by the metric~\eqref{rm}. Since $g_{\phi\phi}\propto F^2$, this means that it is possible for $g_{\phi\phi}$ to have both signs and for the metric to have closed timelike curves (CTCs) without harming the finiteness of $\mathcal{R}$.

  \item The Kretschmann scalar is also finite everywhere
\begin{equation*}
    R_{\al\bt\mu\nu}R^{\al\bt\mu\nu}=P_{K}/(r^2\ro^{12}A^4),
\end{equation*}
where $P_K$ is a polynomial in the same functions, variables, and parameters on which $P_C$ is dependent.

  \item The only nonvanishing components of the Einstein tensor corresponding to~\eqref{rm} are $G_{tt}$, $G_{rr}$, $G_{\ta\ta}$, $G_{t\phi}$, and $G_{\phi\phi}$. Its other components are identically zero.

  \item The flare-out condition is derived taking the derivative of~\eqref{p1} wrt $r$:
\begin{equation}\label{p2}
\frac{B-rB'}{r^2}= \frac{\De}{\ro^2 A}\,\frac{b-rb'}{r^2}+\Big(\frac{\De}{\ro^2 A}\Big)'(1-b/r).
\end{equation}
Near the throat the second term approaches zero. Since $\De/(\ro^2 A)>0$, the fourth line~\eqref{g2} implies
\begin{equation}
rB'<B\;(\text{near the throat}),
\end{equation}
which is the same as the flare-out condition for a nonrotating wormhole [fourth line~\eqref{g2}]. Using the fifth line~\eqref{g2} along with $B(r_0,\ta_0)=r_0$ and $b(r_0)=r_0$ in~\eqref{p2}, we obtain
\begin{equation}
B'(r_0,\ta_0)\leq 1.
\end{equation}

  \item The asymptotical flatness of the nonrotating wormhole, as defined in the first and third lines~\eqref{g2}, ensures that of the rotating wormhole~\eqref{rm}:
      \begin{equation}
        \lim_{r\to\infty}g_{tt}=1\quad\text{ and }\quad \lim_{r\to\infty}B/r=0.
      \end{equation}

  \item If the mass of the nonrotating wormhole is finite, Eq.~\eqref{g3} holds. This latter along with $\lim_{r\to\infty}A=1$ yield
      \begin{equation}
        \lim_{r\to\infty}B=2M,
      \end{equation}
      which we take as twice the mass of the rotating wormhole.

  \item As in the case of the Kerr solution, it is straightforward to show that the zero-angular-momentum observers (ZAMOs) possess an angular velocity equal to that of the rotating wormhole $\om$~\eqref{rm2}.

  \item Since the nonrotating wormhole~\eqref{g1} has no horizon, $A$ is never 0: $A>0$. This implies that $g_{tt}=\De_{\ta}/\ro^2>0$~\eqref{rm3}; thus, the rotating wormhole~\eqref{rm} does not develop an ergosphere region around the throat.

  \item The flattening coefficient $F^2$, $\Si$, and $g_{\phi\phi}$ are all proportional to $r^4+a^4\cos^2\ta+a^2r^2(2-A\sin^2\ta)$. If $A>2$, $g_{\phi\phi}$ may turn negative for some value(s) ($r_1,\ta_1$) of ($r,\ta$). Since $A$ has to approach 1 at spatial infinity~\eqref{g2}, this change in the sign of $g_{\phi\phi}$ may occur only near the throat. If this is the case (i.e. if $r_1>r_0$), the rotating wormhole develops CTCs near the throat, for $\phi$ becomes a timelike coordinate. From now on we only consider nonrotating wormholes with
\begin{equation}\label{p3}
    A(r)\leq 2\quad\text{for all }\;r,
\end{equation}
so that their rotating counterparts do not develop CTCs.

  \item With the restriction~\eqref{p3}, the flattening coefficient $F^2=\Si/\ro^4$ is always greater than, or equal, 1
  \begin{equation}
   F^2-1=\frac{(2-A)r^2+a^2\cos^2\ta}{(r^2+a^2\cos^2\ta)^2}\,a^2\sin^2\ta\geq 0.
  \end{equation}
\end{itemize}

In the following sections we shall use the metric~\eqref{rm} to generate one-, two-, and three-fluid-sourced rotating wormholes along with their nonrotating counterparts. Given any static wormhole solution ($A,b$)~\eqref{g1} to the field equations $G^{\mu\nu}=8\pi T^{\mu\nu}$ of general relativity or to the equations $G^{\mu\nu}=8\pi T_{\text{eff}}^{\mu\nu}$ of any generalized theory of gravitation ($T_{\text{eff}}$ being the effective SET), it suffices to inject it in~\eqref{rm} to get its rotating counterpart. However, the purpose of this work is to focus on, and derive, fluid-sourced wormholes; that is, solutions generated by anisotropic fluids in motion. We seek solutions endowed with interesting physical and geometrical properties, which will result in imposing in each case a formula constraining $A$ and $b$.

\section{One-fluid-sourced rotating wormholes\label{secof}}

We choose a reference frame $(e_t,\,e_r,\,e_{\ta},\,e_{\phi})$ dual to the 1-forms defined in~\eqref{rm1}, $\om^t\equiv \sqrt{\De/\ro^2}(\dd t-a\sin^2\ta\dd \phi)$, $\om^r\equiv -\sqrt{\ro^2/\De}\sqrt{rA/(r-b)}\dd r$, $\om^{\ta}\equiv -\sqrt{\ro^2}\dd \ta$, $\om^{\phi}\equiv (\sin\ta/\sqrt{\ro^2}) [a\dd t-(r^2+a^2)\dd \phi]$:
\begin{align}
&e^{\mu}_t=\frac{(r^2+a^2,0,0,a)}{\sqrt{\ro^2\De}},
& & e^{\mu}_r=\sqrt{\frac{\De}{\ro^2}}\sqrt{\frac{r-b}{rA}}\,(0,1,0,0),\nn\\
\label{of1}&e^{\mu}_{\ta}=\frac{(0,0,1,0)}{\sqrt{\ro^2}},
& & e^{\mu}_{\phi}=\frac{(a\sin^2\ta,0,0,1)}{\sqrt{\ro^2}\sin\ta}.
\end{align}
The source term in the field equations is taken as an anisotropic fluid whose SET is of the form
\begin{equation}\label{of2}
    T^{\mu\nu}=\ep e^{\mu}_te^{\nu}_t+p_re^{\mu}_re^{\nu}_r+p_{\ta}e^{\mu}_{\ta}e^{\nu}_{\ta}+p_{\phi}e^{\mu}_{\phi}e^{\nu}_{\phi},
\end{equation}
where we use the same notation ($\ep,p_r$) as for the nonrotating wormhole~\eqref{g3b} but the values of ($\ep,p_r$) are generally different from their nonrotating counterparts. The transverse pressure is not isotropic in the rotating case and splits into two components ($p_{\ta},p_{\phi}$). The SET can be brought to the standard form in arbitrary coordinates
\begin{equation}\label{of3}
T_{\mu\nu}=(\ep+p)u_{\mu}u_{\nu}-pg_{\mu\nu}+\Pi_{\mu\nu},
\end{equation}
where $u^{\mu}$ is the 4-velocity vector of the fluid and $\ep=u^{\mu}u^{\nu}T_{\mu\nu}$
\begin{equation}\label{of3a}
    u^{\mu}=\frac{e^{\mu}_t+U_1e^{\mu}_r+U_2e^{\mu}_{\ta}+U_3e^{\mu}_{\phi}}{\sqrt{1-U_1^2-U_2^2-U_3^2}}.
\end{equation}
$\Pi_{\mu\nu}$ is the traceless anisotropic pressure tensor and $p$ is the average isotropic pressure defined in terms of the orthogonal projector $h^{\mu}{}_{\nu}=\de^{\mu}{}_{\nu}-u^{\mu}u_{\nu}$ on $u^{\mu}$ by
\begin{equation}
\Pi_{\mu\nu}=\Pi(s_{\mu}s_{\nu}+\tfrac{1}{3}h_{\mu\nu}),\quad p=-\frac{h^{\mu\nu}T_{\mu\nu}}{3},
\end{equation}
where $s^{\mu}$ is a unit spacelike 4-vector orthogonal to $u^{\mu}$: $u^{\mu}s_{\mu}=0$. $s^{\mu}$ is proportional to $S_1e^{\mu}_t+e^{\mu}_r+S_2e^{\mu}_{\ta}+S_3e^{\mu}_{\phi}$ but, without loss of generality, we can take $S_2=S_3=0$ leaving $s^{\mu}$ of the form
\begin{equation}\label{of3b}
    s^{\mu}=\frac{U_1e^{\mu}_t+e^{\mu}_r}{\sqrt{1-U_1^2}}.
\end{equation}
If $p_s=s^{\mu}s^{\nu}T_{\mu\nu}$ denotes the pressure along $s^{\mu}$ then
\begin{equation}
\Pi=\tfrac{3}{2}(p_s-p)=p_s-p_t,
\end{equation}
where $p_t$ is the average isotropic transverse pressure defined in terms of $t^{\mu}$ (a unit spacelike 4-vector orthogonal to $u^{\mu}$ and $s^{\mu}$) by: $p_t=t^{\mu}t^{\nu}T_{\mu\nu}$.

Another useful expression for $T_{\mu\nu}$ is~\cite{MH}
\begin{equation}
T_{\mu\nu}=(\ep+p_t)u_{\mu}u_{\nu}-p_tg_{\mu\nu}+(p_s-p_t)s_{\mu}s_{\nu}.
\end{equation}

In the case of~\eqref{of2}, if we assume $u^{\mu}$ along $e^{\mu}_t$, we obtain $u^{\mu}= e^{\mu}_t$, $s^{\mu}= e^{\mu}_r$, $p_s=p_r$, $p= (p_r+p_{\ta}+p_{\phi})/3$, and $p_t= (p_{\ta}+p_{\phi})/2$. Assuming $u^{\mu}= e^{\mu}_t$ we infer that the fluid elements rotate with the differential angular velocity $\Om=\dd \phi/\dd t=a/(r^2+a^2)$~\eqref{of1} that is different from that of the rotating wormhole $\om$~\eqref{rm2}. The fluid elements do not follow geodesic motion.

The nonvanishing components of the SET~\eqref{of2} are given by the matrix
\begin{equation}\label{of4a}
T^{\mu}{}_{\nu }=\begin{bmatrix}
 \frac{r^2\epsilon  +a^2 (\epsilon +p_{\phi } \sin ^2\theta )}{\ro^2} & 0 & 0 & -\frac{a (a^2+r^2) (\epsilon
+p_{\phi }) \sin ^2\theta }{\ro^2} \\
 0 & -p_r & 0 & 0 \\
 0 & 0 & -p_{\theta } & 0 \\
 \frac{a (\epsilon +p_{\phi })}{\ro^2} & 0 & 0 & -\frac{r^2p_{\phi }+a^2 (p_{\phi }+\epsilon  \sin ^2\theta )}{\ro^2}
\end{bmatrix}.
\end{equation}
The separate resolutions of the field equations $G^{r}{}_{r}=8\pi T^{r}{}_{r}$ and $G^{\ta}{}_{\ta}=8\pi T^{\ta}{}_{\ta}$ provide the expressions of ($p_r,p_{\ta}$) in terms of ($A,b,r,\ta,a$). Similarly, the resolution of the system $\{G^{t}{}_{t}=8\pi T^{t}{}_{t},G^{\phi}{}_{\phi}=8\pi T^{\phi}{}_{\phi}\}$ provides expressions for ($\ep,p_{\phi}$) which satisfy the field equation $G_{t\phi}=8\pi T_{t\phi}$ only if we constrain ($A,b$) by
\begin{equation}\label{of4}
r (r-b) A'+[b-r (2-b')]A+2 r A^2=0.
\end{equation}
Since $A$ cannot be zero on the throat $r_0=b(r_0)$, Eq.~\eqref{of4} implies
\begin{equation}\label{of5}
A(r_0)=\frac{1-b'(r_0)}{2}.
\end{equation}
Solving~\eqref{of4} with the initial condition~\eqref{of5} we obtain
\begin{equation}\label{of6}
    A=\frac{r (r-b)}{r^2-r_0^2}\quad\text{or}\quad b=r-(r^2-r_0^2)\frac{A}{r}.
\end{equation}
It is easy to show that ($A,b$) as given by~\eqref{of5} and~\eqref{of6} satisfy all the requirements~\eqref{g2} of a nonrotating wormhole. This will be shown soon later.

The final expression of the SET is given by
\begin{align}
\label{set1}&\epsilon =\frac{r^2 b'}{8 \pi  \ro^4}+\frac{a^2 r_0^2 [(3A-1)\cos ^2\theta-2]}{8 \pi  \ro^6},\\
&p_r=-\epsilon -\frac{r_0^2 \De}{4 \pi  \ro^6},\nn
\end{align}
\begin{align}
&p_{\ta}=\frac{2A^2+(r A'-2) A+(r-b) (3 A'+r A'')}{16 \pi  \rho ^2 A} -p_r,\nn\\
\label{set4}&p_{\phi }=p_{\theta }+\frac{a^2 r_0^2 \sin ^2\theta }{4 \pi  \ro^6},
\end{align}
where~\eqref{of4} and~\eqref{of6} were used to reduce the expression of the SET. The expression of $\ep$ manifestly generalizes the first line~\eqref{g3b} to the rotating case. The expressions of $p_r$ and $p_{\theta }+p_{\phi}$ could be arranged in the following forms
\begin{align*}
(\ln  A)'& =\frac{8 \pi  \ro^6 p_r+r (\ro^2+ a^2 \cos ^2\theta ) b}{r^2 \ro^2(r-b)}\\
\quad & +\frac{a^2 \cos ^2\theta }{r \ro^2 A}-\frac{(2-A) a^2 \cos ^2\theta }{\ro^2 (r-b)},\\
p_{\theta }+p_{\phi }& =\frac{[\Delta +\rho ^2 A] p_r}{\Delta }+\frac{\rho ^2 p_r{}'}{r}+\frac{r\rho ^2 (\epsilon +p_r) A'}{2 \Delta }\\
\quad & -\frac{a^2 (1-A\cos ^2\theta) \epsilon }{\Delta },
\end{align*}
which manifestly generalize the second and third lines~\eqref{g3b}.

Next, we show that ($A,b$) as given by~\eqref{of5} and~\eqref{of6} satisfy all the requirements~\eqref{g2} of a nonrotating wormhole. Eq.~\eqref{of6} implies $b(r_0)=r_0$. Using~\eqref{of6} in the second line~\eqref{g2} we obtain $(r^2-r_0^2)A\geq 0$, which is always satisfied ($A>0$). Eq.~\eqref{of5} implies $b'(r_0)=1-2A(r_0)<1$. The derivative of $b$~\eqref{of6} reduces to
\begin{align}
rb'& = r(1-A)-\frac{r_0^2}{r}A-(r^2-r_0^2)A'\nn\\
\label{req}\quad & = b-\frac{2r_0^2}{r}A-(r^2-r_0^2)A'.
\end{align}
Near the throat, the term $(r^2-r_0^2)A'$ is neglected with respect to the other terms yielding $rb'\simeq b-(2r_0^2/r)A<b$ [fourth line~\eqref{g2}]. Finally, for the nonrotating wormhole given by~\eqref{of6}, it is straightforward to see that the line 1 of~\eqref{g2} ($\lim_{r\to\infty}A=1$) implies its line 3 ($\lim_{r\to\infty}b/r=0$) and conversely.

Now, let us see under which condition the rotating counterparts of the family of static solutions~\eqref{of6} do not develop CTCs~\eqref{p3} near the throat. Using~\eqref{of5}, the constraint $A(r_0)< 2$ yields
\begin{equation}\label{ctc1}
b'(r_0)>-3,
\end{equation}
while the constraint $A(r)< 2$ for $r>r_0$ implies
\begin{equation*}
r(r+b)-2r_0^2\geq 0\qquad (r>r_0).
\end{equation*}
To ensure positiveness of the lhs for all $r> r_0$, its derivative must be positive yielding the further constraint on the value of $b'(r)$
\begin{equation}\label{ctc2}
2r+b+rb'>0\qquad (r\geq r_0),
\end{equation}
which holds for $r=r_0$ too~\eqref{ctc1}. Similarly, the constraint $A< 1$ is satisfied if
\begin{equation}\label{ctc3}
b+rb'>0\qquad (r\geq r_0).
\end{equation}

From now on, we only consider massive nonrotating wormholes having positive energy densities $\ep=b'/(8\pi r^2)\geq 0$. We choose this type of wormholes because they violate the least the local energy conditions. Since $b'(r)\geq 0$ for all $r$, the constraints~\eqref{ctc1} and~\eqref{ctc2} are satisfied and thus no closed timelike curve exists near the throat of the corresponding rotating wormholes. Moreover, since~\eqref{ctc3} is also satisfied, $A$ remains smaller than unity.

Instances of such nonrotating wormholes are the massive solutions with mass $M$ given by
\begin{equation}\label{of7}
b=2M,\quad A=1-\frac{2M}{r+2M},\quad 2M=r_0,
\end{equation}
\begin{equation}\label{of8}
\hspace{-1mm}b=2M-\frac{(2M-r_0)r_0}{r},\;\; A=1-\frac{2M}{r+r_0},\;\; M<r_0.
\end{equation}
In both instances $A$ is an increasing function of $r$ and bounded from above by 1. In~\eqref{of8}, the constraint $M<r_0$ is to have $b'(r_0)<1$~\eqref{g2} and $A(r_0)>0$ ensuring positiveness of $A$ for all $r$. The corresponding SETs are derived from Eqs.~\eqref{set1}-\eqref{set4} on setting $a^2=0$.

It is straightforward to generalize the above expressions of ($A,b$). In terms of the dimensionless variables
\begin{equation}\label{of8b}
    y\equiv r/r_0,\quad m\equiv M/r_0,
\end{equation}
we obtain the general solution
\begin{align}
\label{of9}&\frac{b}{r_0}=2m-\frac{(2m-1)}{y^{\bt}},\quad \bt >0, \quad \tfrac{1}{2}<m<\tfrac{1+\bt}{2\bt},\\
\label{of10}&A=y^{1-\bt}\Big(\frac{y^{1+\bt}-1}{y^2-1}-2m\frac{y^{\bt}-1}{y^2-1}\Big).
\end{align}
Here again the constraint $m<(1+\bt)/(2\bt)$ is to have $b'(r_0)<1$ and $b<r$ for $r>r_0$~\eqref{g2}, and to ensure positiveness of $A$ for all $r\geq r_0$. Notice that the constraint $m<(1+\bt)/(2\bt)$  not only ensures $b'(r_0)<1$ but $b'(r)<1$ for all $r\geq r_0$ as well.

Nonrotating wormholes with $\ep=b'/(8\pi r^2)\geq 0$ have the property that $m\geq 1/2$~\cite{types}. In the special case of the solutions~\eqref{of9}, this reduces to $m>1/2$. The corresponding SET is derived from Eqs.~\eqref{set1}-\eqref{set4} on setting $a^2=0$.

The dragging effects of the rotating wormholes given by Eqs.~\eqref{rm}, \eqref{of9}, \eqref{of10}, and Eqs.~\eqref{set1}-\eqref{set4} are more pronounced for $\bt<1$ than for $\bt>1$. At spatial infinity, this becomes obvious if we expand the angular velocity $\om$~\eqref{rm2} in powers of $1/r$
\begin{equation}\label{of11}
    r_0^2\om=\left\{
               \begin{array}{ll}
                 a\big(\frac{2m}{y^3}-\frac{2m-1}{y^{3+\bt}}-\frac{1}{y^4}+\cdots\big), & 0<\bt\leq 1; \\
                 a\big(\frac{2m}{y^3}-\frac{1}{y^4}+\cdots\big), & \bt>1,
               \end{array}
             \right.
\end{equation}
where the leading term of $\om$, $2aM/r^3$, is that of an ordinary star. This is different from the corresponding expression for the Kerr metric where the term in $1/r^4$ is absent. The measurements of the dragging effects, far away from the sources, allow to distinguish these rotating wormholes from the Kerr black hole.

Since $\om$ is an increasing function of $\sin^2\ta$, the dragging effects do depend on $\ta$. For a fixed value of the radial coordinate, the dragging effects are more accentuated in the equatorial plane than elsewhere. This property is not specific to the rotating wormholes given by Eqs.~\eqref{rm}, \eqref{of9}, \eqref{of10}, and Eqs.~\eqref{set1}-\eqref{set4}; rather, it applies to all rotating wormholes given by Eq.~\eqref{rm} since the only dependence on $\sin^2\ta$ occurs in $\Si$~\eqref{rm} which has the same shape for all solutions.

For these one-fluid sourced massive rotating wormholes (with positive energy density at least in the nonrotating case) we see that there is no way to reduce the scope of the dragging effects to that of ordinary stars $\om\sim 2aM/r^3$; the two-fluid sourced rotating wormholes, derived by the superposition of fields~\cite{Trot}, have offered such possibilities.

To the form~\eqref{rm2} of the rotating metric are associated the set of 1-forms $\bar{\om}^t\equiv \sqrt{\ro^2\De/\Si}\dd t$, $\bar{\om}^r\equiv \om^r = -\sqrt{\ro^2/\De}\sqrt{rA/(r-b)}\dd r$, $\bar{\om}^{\ta}\equiv \om^{\ta} = -\sqrt{\ro^2}\dd \ta$, $\bar{\om}^{\phi}\equiv \sqrt{\Si/\ro^2}\sin\ta (\om\dd t-\dd \phi)$ and the corresponding frame
\begin{align}
\label{of12}&\bar{e}^{\mu}_t=\Big(\sqrt{\frac{\Si}{\ro^2\De}},0,0,\frac{2af}{\sqrt{\ro^2\De\Si}}\Big),
& & \bar{e}^{\mu}_r=e^{\mu}_r,\\
&\bar{e}^{\mu}_{\ta}=\e^{\mu}_{\ta},
& & \bar{e}^{\mu}_{\phi}=\Big(0,0,0,\frac{\sqrt{\ro^2}}{\sqrt{\Si}\sin\ta}\Big).\nn
\end{align}
Similarly, to the form~\eqref{rm3} of the rotating metric are associated the set of 1-forms $\tilde{\om}^t\equiv (\De_{\ta}\dd t+2af\sin^2\ta\dd \phi)/\sqrt{\ro^2\De_{\ta}}$, $\tilde{\om}^r\equiv \om^r$, $\tilde{\om}^{\ta}\equiv \om^{\ta}$, $\tilde{\om}^{\phi}\equiv -\sqrt{\ro^2\De/\De_{\ta}}\sin\ta \dd \phi$ and the corresponding frame
\begin{align}
\label{of13}&\tilde{e}^{\mu}_t=\Big(\sqrt{\frac{\ro^2}{\De_{\ta}}},0,0,0\Big),
& & \hspace{-3mm}\tilde{e}^{\mu}_r=e^{\mu}_r,\\
&\tilde{e}^{\mu}_{\ta}=\e^{\mu}_{\ta},
& & \hspace{-3mm}\tilde{e}^{\mu}_{\phi}=\Big(\frac{-2af\sin\ta}{\sqrt{\ro^2\De\De_{\ta}}},0,0,\frac{\sqrt{\De_{\ta}}}{\sqrt{\ro^2\De}\sin\ta}\Big).\nn
\end{align}
The rotating wormholes derived in this section [Eqs.~\eqref{rm}, \eqref{of9}, \eqref{of10}, and Eqs.~\eqref{set1}-\eqref{set4}] have been determined using the frame~\eqref{of1} to expand the SET. Had we used either the frame~\eqref{of12} or~\eqref{of13} we would have found other rotating wormhole solutions. However, if we restrict ourselves to solutions where $b$ is given by~\eqref{of9} (without the constraint $m>1/2$) and we use either the frame~\eqref{of12} or~\eqref{of13}, we can show that the only existing solution is the Schwarzschild wormhole
\begin{equation}
b=2M=r_0,\quad A=1-\frac{2M}{r}\qquad (m=1/2),
\end{equation}
so that the corresponding rotating solution is the ``Kerr wormhole".

\section{Two-fluid-sourced rotating wormholes\label{sectf}}

We intend to determine two-fluid sourced rotating wormholes by the method of superposition of fields~\cite{snk,ex,az3,BK,Trot}. This will allow us to construct rotating wormholes endowed with interesting physical properties. Applied to derive static solution, the method consists in splitting the SET into a sum of sub-SETs~\cite{snk,ex,BK}. This way of splitting still applies to rotating solutions~\cite{az3}; however, in this case it is possible to generalize the method, as we did in~\cite{Trot}, by attaching to each selected moving (here rotating) frame a form of matter, that is, a sub-SET $T^{\mu\nu}$.

In the case of two-fluid-sourced rotating wormholes, the generalized method consists in splitting the total SET as $T^{\mu\nu}+\bar{T}^{\mu\nu}$, $T^{\mu\nu}+\tilde{T}^{\mu\nu}$, or $\bar{T}^{\mu\nu}+\tilde{T}^{\mu\nu}$ where each component $T^{\mu\nu}$, $\bar{T}^{\mu\nu}$, and $\tilde{T}^{\mu\nu}$ is expanded, as in~\eqref{of2}, using the frame~\eqref{of1}, \eqref{of12}, and~\eqref{of13}, respectively, with
\begin{align}
\label{tf1}&\bar{T}^{\mu\nu}=\bar{\ep} \bar{e}^{\mu}_t\bar{e}^{\nu}_t+\bar{p}_r\bar{e}^{\mu}_r\bar{e}^{\nu}_r+\bar{p}_{\ta}\bar{e}^{\mu}_{\ta}\bar{e}^{\nu}_{\ta}
+\bar{p}_{\phi}\bar{e}^{\mu}_{\phi}\bar{e}^{\nu}_{\phi},\\
\label{tf2}&\tilde{T}^{\mu\nu}=\tilde{\ep} \tilde{e}^{\mu}_t\tilde{e}^{\nu}_t+\tilde{p}_r\tilde{e}^{\mu}_r\tilde{e}^{\nu}_r+
\tilde{p}_{\ta}\tilde{e}^{\mu}_{\ta}\tilde{e}^{\nu}_{\ta}+\tilde{p}_{\phi}\tilde{e}^{\mu}_{\phi}\tilde{e}^{\nu}_{\phi}.
\end{align}

We start with the case where the SET is $T^{\mu\nu}+\bar{T}^{\mu\nu}$. Here $T^{\mu\nu}$ is given by~\eqref{of4a} and $\bar{T}^{\mu\nu}$~\eqref{tf1} reads
\begin{equation}\label{tf3}
\bar{T}^{\mu}{}_{\nu }=\begin{bmatrix}
 \bar{\ep} & 0 & 0 & 0 \\
 0 & -\bar{p}_r & 0 & 0 \\
 0 & 0 & -\bar{p}_{\theta } & 0 \\
 \frac{2a f(\bar{\ep} +\bar{p}_{\phi })}{\Si} & 0 & 0 & -\bar{p}_{\phi }
\end{bmatrix}.
\end{equation}
Notice that the number of unknowns (the eight components of $T^{\mu\nu}$ and $\bar{T}^{\mu\nu}$ and $A$ and $b$) exceeds the number of the field equations [$G_{\mu\nu}=8\pi (T_{\mu\nu}+\bar{T}_{\mu\nu})$], which is five. This is the advantage of the method of superposition of fields, for this will allow us to fix the values of some unknowns to well-defined physical entities and will yield interesting physical rotating wormholes. For instance, we assume that the SET $\bar{T}_{\mu\nu}$ is of electromagnetic nature. Taking into account the nature of the field equations, which are split into two \emph{independent} sets
\begin{align}
\label{tf4}\hspace{-9mm}&\text{S1:}\; G^{r}{}_{r}=8\pi (T^{r}{}_{r}+\bar{T}^{r}{}_{r}), \; G^{\ta}{}_{\ta}=8\pi (T^{\ta}{}_{\ta}+\bar{T}^{\ta}{}_{\ta}),\\
\hspace{-9mm}&\text{S2:}\;G^{t}{}_{t}=8\pi (T^{t}{}_{t}+\bar{T}^{t}{}_{t}),\;G^{\phi}{}_{\phi}=8\pi (T^{\phi}{}_{\phi}+\bar{T}^{\phi}{}_{\phi}),\nn\\
\label{tf5}& \quad\;\; G_{t\phi}=8\pi (T_{t\phi}+\bar{T}_{t\phi}),
\end{align}
we may fix the values of ($\bar{\ep},\bar{p}_r,\bar{p}_{\ta}$) to
\begin{equation}\label{tf6}
   \bar{\ep}=-\bar{p}_r=\bar{p}_{\ta}\equiv \frac{Q^2}{8\pi\ro^4},
\end{equation}
as in the Kerr solution, but not the value of $\bar{p}_{\phi}$ which is determined upon solving the set S2. The resolution of the latter provides unique values for ($\ep,p_{\phi},\bar{p}_{\phi}$). These expressions are sizeable but could be simplified noticing that the limit of $\bar{p}_{\phi}$ as $a^2\to 0$, which is given by
\begin{equation}
   \lim_{a^2\to 0}\bar{p}_{\phi}=\frac{r^2 (r-b) A'+r[b-r (2-b')]A+2 (r^2-Q^2) A^2}{16\pi r^4A^2},
\end{equation}
should reduce to the static value $Q^2/(8\pi r^4)$. This yields the following relation between $A$ and $b$
\begin{equation}\label{tf7}
r^2 (r-b) A'+r[b-r (2-b')]A+2 (r^2-2Q^2) A^2=0.
\end{equation}
This generalizes~\eqref{of4}, which applies to the case $\bar{T}_{\mu\nu}\equiv 0$, to solutions where $Q\neq 0$. Since $A>0$, Eq.~\eqref{tf7} yields
\begin{equation}\label{tf8}
\hspace{-4mm}A(r_0)=\left\{
               \begin{array}{ll}
                 \dfrac{r_0^2[1-b'(r_0)]}{2(r_0^2-2Q^2)}, & r_0^2>2Q^2\text{ and }b'(r_0)<1; \\
                 \\
                 -\dfrac{r_0b''(r_0)}{4}, &  r_0^2=2Q^2\text{ and }b'(r_0)=1.
               \end{array}
             \right.
\end{equation}
By the fourth line~\eqref{g2}, $1-b'(r_0)$ cannot be negative, so is the term $r_0^2-2Q^2$ in the first case~\eqref{tf8} [otherwise, $A(r_0)$ would be negative]. This sets an upper limit for the value of the charge for these nonrotating, and their rotating counterpart, wormholes
\begin{equation}\label{tf9}
    q^2\leq 1/2\qquad (q\equiv Q/r_0).
\end{equation}
Equation~\eqref{tf7} can be solved for either $A$ or $b$ yielding
\begin{align}
\label{tf10}&A=\frac{r (r-b)}{r^2-r_0^2-4 Q^2 \ln (r/r_0)},\\
\label{tf11}&b=r-(r^2-r_0^2) \frac{A}{r}+\frac{4 Q^2 A \ln (r/r_0)}{r},
\end{align}
where we have imposed the condition $b(r_0)=r_0$ to fix the constant of integration. We see that the constraint~\eqref{tf9} ($r\geq r_0$) ensures that the denominator in~\eqref{tf10} is monotonically increasing function of $r$ keeping $A>0$ for all $r$.

By a similar discussion to the one given in the paragraph containing~\eqref{req} one can show that ($A,b$) as given by~\eqref{tf10} and~\eqref{tf11} satisfy all the requirements~\eqref{g2} of a nonrotating wormhole. For instance, $rb'$ takes the form
\begin{multline}
rb'=b-2\frac{r_0^2-2Q^2+4 Q^2 \ln (r/r_0)}{r}A\\-[r^2-r_0^2-4 Q^2 \ln (r/r_0)]A'.
\end{multline}
implying $rb'<b$ near the throat.

It is easy to establish that the rotating counterparts of the family of static solutions~\eqref{tf10} do not develop CTCs~\eqref{p3} \emph{near} the throat if
\begin{align}
&q^2<\frac{b'(r_0)+3}{8}<\frac{1}{2},\quad b'(r_0)<1,\nn\\
\label{ctc4}&r_0b''(r_0)>-5-3b'(r_0),\quad\text{or},\\
\label{ctc5}&q^2=\frac{1}{2},\quad b'(r_0)=1,\quad r_0b''(r_0)>-8.
\end{align}
These provide stronger constraints than~\eqref{tf9}.

From~\eqref{tf10} we see that $1/(r-b)$ and $Ar/(r-b)$ are functions of ($A,Q^2,r_0$) only. Thus, the nonrotating~\eqref{g1} and rotating~\eqref{rm} wormholes are expressed explicitly in terms of ($A,Q^2,r_0$) only, and any dependence on the mass $M$ is incorporated in $A$. The rotating wormhole takes the final expression
\begin{align}
&\dd s^2 =\Big(1-\frac{2f}{\ro^2}\Big)\dd t^2-\frac{r^2\ro^2 \dd r^2}{\De [r^2-r_0^2-4 Q^2 \ln (r/r_0)]}\nn\\
\label{rmq}&+\frac{4af \sin ^2\theta}{\ro^2}\,\dd t\dd \phi-\ro^2\dd \ta^2-\frac{\Si\sin ^2\theta}{\ro^2}\,\dd \phi^2,\\
&\ro^2\equiv r^2+a^2 \cos^2\ta,\quad 2f(r)\equiv r^2(1-A)\nn\\
&\De(r)\equiv r^2 A+a^2,\quad \Si\equiv (r^2+a^2)^2-a^2\De\sin^2\ta \nn.
\end{align}
The nonrotating metric is derived setting $a^2=0$. If $Q^2\equiv 0$, we reobtain the nonrotating and rotating metrics derived in the previous section, which were not written explicitly [they are special cases of~\eqref{rmq}]. Eq.~\eqref{rmq} constitutes a family of rotating and nonrotating solutions where $A$ and $b$ are related by~\eqref{tf8}, \eqref{tf10}, and~\eqref{tf11}. If $Q^2\equiv 0$, the family of solutions is sourced by an exotic fluid given by Eqs.~\eqref{of5}-\eqref{set4}. If $Q^2\neq 0$, the family of solutions is sourced by two fluids, one of which, $\bar{T}_{\mu\nu}$, is electromagnetic given by~\eqref{tf6} and\footnote{The rhs of~\eqref{tf12} is
\begin{equation*}
\frac{Q^2 [r^4+3 a^2 r^2+2 a^2 (a^2 \cos ^2\theta -r^2 A \sin ^2\theta )]}
{8 \pi  r^2 (r^2+a^2) \rho ^4},
\end{equation*}
which reduces to the static value $Q^2/(8\pi r^4)$ if rotation is suppressed setting $a^2=0$.}
\begin{equation}\label{tf12}
\bar{p}_{\phi}=\frac{Q^2 [2\Si -r^2(r^2+a^2)]}
{8 \pi  r^2 (r^2+a^2) \rho ^4}.
\end{equation}
The other fluid, $T_{\mu\nu}$, is exotic given by
\begin{align}\label{tf13}
&p_r=\frac{r}{8\pi\rho ^6 A} \big\{[2 a^2 r \cos ^2\theta-(\rho ^2+a^2 \cos ^2\theta ) b] A\\
&-a^2 r A^2 \cos ^2\theta +(r-b) (r \rho ^2A'-a^2 \cos ^2\theta )\big\}+\frac{Q^2}{8\pi \ro^4},\nn
\end{align}
\begin{align}\label{tf14}
&p_{\ta}=\frac{1}{16 \pi  r^2 \rho ^2 A} \big\{2 (r^2-2 Q^2) A^2+r [(r^2-2 Q^2) A'-2 r] A\nn\\
&+r^2 (r-b) (3 A'+r A'')\big\}-p_r,
\end{align}
\begin{align}\label{tf15}
&p_{\phi} =p_{\ta}+\frac{a^2 r_0^2 \sin ^2\theta }{4 \pi  \ro^6}-\frac{Q^2}{8\pi \ro^4}\nn\\
&+\frac{Q^2}{8 \pi  r^2 \rho ^6}\big\{a^2 r^2 [8 \ln(r/r_0) \sin ^2\theta -(2+\cos ^2\theta )]\nn\\
&-(r^4+2 a^4 \cos ^2\theta )\big\},
\end{align}
\begin{align}\label{tf16}
&\ep =\frac{1}{8 \pi  \rho^6}\big\{[2 r^2 (Q^2-r^2)+a^2 (2 Q^2+r^2) \cos ^2\theta ] A\nn\\
&-r (r^2-2 a^2 \cos ^2\theta ) b\big\}+\frac{a^2 r (2+\cos ^2\theta ) (r-b)}{8 \pi  \rho ^6 A}\nn\\
&-\frac{r^2 (r-b) A'}{8 \pi  \rho ^4 A}+\frac{Q^2 (2
a^2-r^2)+2 r^2 (r^2-a^2)}{8 \pi  \rho ^6}\\
&+\frac{a^2 [Q^2 (2 a^2-r^2)-2 r^4] \cos ^2\theta }{8 \pi  r^2 \rho
^6}+\frac{Q^2 \Delta }{4 \pi  r^2 (a^2+r^2) \rho ^2},\nn
\end{align}
where~\eqref{tf7} has been used to eliminate $b'$ from the expressions of ($\ep,p_{\ta},p_{\phi}$) and~\eqref{tf11} has been used to eliminate $b$ from the expression of $p_{\phi}$.

We keep working with $b$ of the form~\eqref{of9} yielding
\begin{equation}\label{tf17}
\hspace{-3mm}A=y^{1-\bt}\Big(\frac{y^{1+\bt}-1}{y^2-1-4q^2\ln y}-2m\frac{y^{\bt}-1}{y^2-1-4q^2\ln y}\Big),
\end{equation}
where $q^2$ is constrained by~\eqref{ctc4} or~\eqref{ctc5} if no CTCs develop near the throat, otherwise $q^2\leq 1/2$~\eqref{tf9}. This new expression of $A$ affects the expansion~\eqref{of11} of the angular velocity $\om$, which now reads
\begin{equation}\label{tf18}
   r_0^2\om=\left\{
               \begin{array}{ll}
                 a\big(\frac{2m}{y^3}-\frac{2m-1}{y^{3+\bt}}-\frac{4q^2\ln y}{y^4}-\frac{1}{y^4}+\cdots\big), &
                 0<\bt<1;\\
                 a\big(\frac{2m}{y^3}-\frac{4q^2\ln y}{y^4}-\frac{2m}{y^4}+\cdots\big), &
                 \bt=1;\\
                 a\big(\frac{2m}{y^3}-\frac{4q^2\ln y}{y^4}-\frac{1}{y^4}+\cdots\big), & \bt>1.
               \end{array}
             \right.
\end{equation}
This is very different from the corresponding expression for the Kerr-Newman metric where the term in $\ln r/r^4$ is absent. For $\bt\geq 1$, we see that the deviation of $\om$ from the corresponding value for ordinary stars, $2aM/r^3$, increases with $q^2$ while for $0<\bt<1$ the effects of the charge take place for much lower values of the radial distance.

The expression of $b$~\eqref{of9} is the simplest form including a finite mass term and yielding a positive energy density. Other expressions, as
\begin{equation*}
    \frac{b}{r_0}=2m-\frac{c}{y^{\bt}}-\frac{2m-1-c}{y^{\gamma}},
\end{equation*}
where $\bt>0,\gamma>0,2m-1>c>0$, are possible but the conclusions drawn in the previous paragraph remain unchanged.

In this section we have treated the case where the SET is the sum $T^{\mu\nu}+\bar{T}^{\mu\nu}$. In the next section we shall treat the case where $\text{SET}=T^{\mu\nu}+\bar{T}^{\mu\nu}+\tilde{T}^{\mu\nu}$ and we shall see that the cases $\text{SET}=T^{\mu\nu}+\bar{T}^{\mu\nu}$ and $\text{SET}=T^{\mu\nu}+\tilde{T}^{\mu\nu}$ are special cases of it. This is why we will skip the case
\begin{equation}\label{skip}
\text{SET}=T^{\mu\nu}+\tilde{T}^{\mu\nu}
\end{equation}
in this section.

There remains the case where the SET is the sum $\bar{T}^{\mu\nu}+\tilde{T}^{\mu\nu}$. This is more involved than the other cases considered in this work, for the equation constraining $A$ and $b$ [compare with~\eqref{tf7}] contains the second derivative of $A$ which makes it impossible to find a closed formula relating $A$ to $b$ as in~\eqref{tf10}
and~\eqref{tf11}.

\section{Three-fluid-sourced rotating wormholes\label{sec3f}}

We intend to determine three-fluid sourced rotating wormholes by the method of superposition of fields. The SET is now the sum of three sub-SETs \[T^{\mu\nu}+\bar{T}^{\mu\nu}+\tilde{T}^{\mu\nu}\] where $T^{\mu\nu}$ and $\bar{T}^{\mu\nu}$ are given by~\eqref{of4a} and~\eqref{tf3} and $\tilde{T}^{\mu\nu}$~\eqref{tf2} reads
\begin{equation}\label{3f1}
\tilde{T}^{\mu}{}_{\nu }=\begin{bmatrix}
 \tilde{\ep} & 0 & 0 & \frac{2a f(\tilde{\ep} +\tilde{p}_{\phi })\sin^2\ta}{\De_{\ta}} \\
 0 & -\tilde{p}_r & 0 & 0 \\
 0 & 0 & -\tilde{p}_{\theta } & 0 \\
 0 & 0 & 0 & -\tilde{p}_{\phi }
\end{bmatrix}.
\end{equation}

The field equations are again split into two independent sets S1 and S2 as in~\eqref{tf4} and~\eqref{tf5} where now the rhs's $T+\bar{T}$ are replaced by the sums $T+\bar{T}+\tilde{T}$. Applying step-by-step the procedure of the previous section, we first fix the values of ($\bar{\ep},\bar{p}_r,\bar{p}_{\ta}$) and ($\tilde{\ep},\tilde{p}_r,\tilde{p}_{\ta},\tilde{p}_{\phi}$) to
\begin{align}
&\label{3f2}\bar{\ep}=-\bar{p}_r=\bar{p}_{\ta}\equiv \frac{Q_1^2}{8\pi\ro^4},\\
&\label{3f3}\tilde{\ep}=-\tilde{p}_r=\tilde{p}_{\ta}=\tilde{p}_{\phi}\equiv \frac{Q_2^2}{8\pi\ro^4},
\end{align}
which correspond to two electromagnetic SETs. The SET $T^{\mu\nu}$ is certainly exotic. The values of ($p_r,p_{\ta}$) are determined upon solving the set S1 and those of ($\ep,p_{\phi},\bar{p}_{\phi}$) are determined upon solving the set S2. In order that $\lim_{a^2\to 0}\bar{p}_{\phi}$ reduces to the static value $Q_1^2/(8\pi r^4)$, we constrain $A$ and $b$ by
\begin{multline}\label{3f4}
r^2 (r-b) A'+\{r[b-r (2-b')]-4Q_2^2\}A\\+2 (r^2-2Q_1^2) A^2=0.
\end{multline}
which generalizes~\eqref{tf7} and reduces to it if the SET $\tilde{T}^{\mu\nu}$ vanishes ($Q_2=0$). In terms of the dimensionless parameters $q_1\equiv Q_1/r_0$ and $q_2\equiv Q_2/r_0$, $A(r_0)$ reads
\begin{equation}\label{3f5}
    A(r_0)=\frac{1-b'(r_0)+4q_2^2}{2(1-2q_1^2)}, \quad (0<q_1^2<1/2).
\end{equation}
Since $1-b'(r_0)\geq 0$ by~\eqref{g2}, this implies that the numerator in the expression of $A(r_0)$ is positive and so is the denominator. Thus, the charge $q_1^2$ is bounded from above by 1/2. With $A(r_0)$ and $b'(r_0)$ being always finite, the case $q_1^2=1/2$ yields a non-wormhole solution, for in this case Eq.~\eqref{3f4} implies $A(r_0)=0$. The fact that $A(r_0)$ is always finite sets an upper limit for $q_2^2$ too. For instance, if we restrict ourselves to rotating wormholes with no CTCs (at least near the throat)~\eqref{p3}, the necessary constraint $A(r_0)\leq 2$ yields on substituting in~\eqref{3f5}
\begin{multline}\label{bnd}
\hspace{-1mm}4q_2^2\leq 3+b'(r_0)-8q_1^2\Rightarrow 0<2q_1^2+q_2^2\leq 1\Rightarrow\\
0<q_1^2<1/2\quad\text{ and }\quad 0<q_2^2<1,
\end{multline}
where we have used $b'(r_0)\leq 1$~\eqref{g2}. Similarly, if we restrict ourselves to rotating wormholes with $A(r)$ increasing and $A(r)<1$, a necessary condition for that is $A(r_0)<1$ yielding
\begin{equation}\label{bnd2}
0<q_1^2+q_2^2\leq \tfrac{1}{2}\Rightarrow 0<q_1^2<\tfrac{1}{2}\;\text{ and }\; 0<q_2^2<\tfrac{1}{2}.
\end{equation}

Equation~\eqref{3f4} can be solved formally for either $A$ or $b$. The expression of the latter reads
\begin{equation}\label{3f6}
    b=r-(r^2-r_0^2) \frac{A}{r}+\frac{4 Q_1^2 A \ln (r/r_0)}{r}+\frac{4Q_2^2A}{r}\int_{r_0}^{r}\frac{\dd u}{uA(u)},
\end{equation}
generalizing~\eqref{tf11}. Using this, the expression of $\bar{p}_{\phi}$ simplifies greatly and generalizes the one given in the previous section~\eqref{tf12}
\begin{equation}\label{3f7}
\hspace{-3mm}\bar{p}_{\phi}=\frac{Q_1^2 [2\Si -r^2(r^2+a^2)]}
{8 \pi  r^2 (r^2+a^2) \rho ^4}+\frac{a^2Q_2^2\Si \cos^2\ta}{4 \pi  r^2 (r^2+a^2)A\De_{\ta} \rho ^4}.
\end{equation}
It is clear that in the limit $a^2\to 0$ we recover the static value $Q_1^2/(8\pi r^4)$. The rotating wormhole takes the final expression
\begin{align}
&\dd s^2 =\Big(1-\frac{2f}{\ro^2}\Big)\dd t^2-\frac{r^2\ro^2 \dd r^2}{\De [r^2-r_0^2-4 Q_1^2 \ln (\frac{r}{r_0})-4 Q_2^2\int_{r_0}^{r}\frac{\dd u}{uA}]}\nn\\
\label{rmq1q2}&+\frac{4af \sin ^2\theta}{\ro^2}\,\dd t\dd \phi-\ro^2\dd \ta^2-\frac{\Si\sin ^2\theta}{\ro^2}\,\dd \phi^2,\\
&\ro^2\equiv r^2+a^2 \cos^2\ta,\quad 2f(r)\equiv r^2(1-A)\nn\\
&\De(r)\equiv r^2 A+a^2,\quad \Si\equiv (r^2+a^2)^2-a^2\De\sin^2\ta \nn.
\end{align}
The nonrotating metric is derived setting $a^2=0$. If $Q_2^2\equiv 0$ and $Q_1^2\equiv Q^2$, we reobtain the nonrotating and rotating wormholes derived in the previous section~\eqref{rmq}. Eq.~\eqref{rmq1q2} constitutes a family of rotating and nonrotating solutions where $A$ and $b$ are related by~\eqref{3f5} and~\eqref{3f6}. If $Q_1^2\equiv 0$ and $Q_2^2\equiv 0$, the family of solutions is sourced by an exotic fluid given by Eqs.~\eqref{of5}-\eqref{set4}. If $Q_1^2\neq 0$ $Q_2^2= 0$, the family of solutions is sourced by two fluids, one of which, $\bar{T}_{\mu\nu}$, is electromagnetic. If $Q_1^2= 0$ and $Q_2^2\neq 0$, the family of solutions is again sourced by two fluids, one of which, $\tilde{T}_{\mu\nu}$, is electromagnetic and the other one, $T_{\mu\nu}$, is exotic. This is the case~\eqref{skip} we skipped in the previous section. Now, if $Q_1^2\neq 0$ and $Q_2^2\neq 0$, the family of solutions is sourced by three fluids, two of which are electromagnetic ($\bar{T}_{\mu\nu},\tilde{T}_{\mu\nu}$) and one is exotic ($T_{\mu\nu}$).

The integral in~\eqref{rmq1q2} could be evaluated closely for a wide choices of $A(r)$. The simplest examples of two-(electrically, magnetically, or both)-charged solutions are (1) the massless wormhole
\begin{align}
&\dd s^2 =\dd t^2-\frac{r^2\ro^2 \dd r^2}{\De [r^2-r_0^2-4 (Q_1^2+Q_2^2) \ln (\frac{r}{r_0})]}\nn\\
\label{rmq1q20}&-\ro^2\dd \ta^2-\frac{\Si\sin ^2\theta}{\ro^2}\,\dd \phi^2,\\
&A\equiv 1. \nn
\end{align}
This could be interpreted as a static, nonrotating, wormhole generated by the three fluids ($T_{\mu\nu},\bar{T}_{\mu\nu},\tilde{T}_{\mu\nu}$), in stationary motion, or as a rotating wormhole with no dragging effects generated by the same three fluids. And (2) the massive rotating wormhole with mass $M<r_0/2$ and metric
\begin{align}
&\dd s^2 =\Big(1-\frac{2f}{\ro^2}\Big)\dd t^2-\ro^2\dd \ta^2\nn\\
&+\frac{4af \sin ^2\theta}{\ro^2}\,\dd t\dd \phi-\frac{\Si\sin ^2\theta}{\ro^2}\,\dd \phi^2\nn\\
\label{rmq1q2m}&-\frac{r^2\ro^2 \dd r^2}{\De [r^2-r_0^2-4 Q_1^2\ln (\frac{r}{r_0})-4 Q_2^2\ln (\frac{r-2M}{r_0-2M})]},\\
&A\equiv 1-\frac{2M}{r},\quad b=2M+Ah(r)>2M,\quad (M<r_0/2),\nn\\
&rh(r)\equiv r_0^2+4 Q_1^2\ln \big(\frac{r}{r_0}\big)+4 Q_2^2\ln \big(\frac{r-2M}{r_0-2M}\big),\nn
\end{align}
generated by the three fluids ($T_{\mu\nu},\bar{T}_{\mu\nu},\tilde{T}_{\mu\nu}$). Both rotating wormholes~\eqref{rmq1q20} and~\eqref{rmq1q2m} do not develop CTCs near the throat.

Now, back to solutions where $b$ is given by~\eqref{of9}. Equation~\eqref{3f6} being not reversible analytically we cannot express $\om$ in terms of $A$. It is however easy to show that the asymptotic behavior of the dragging effects~\eqref{tf18} remains valid at least up to $\ln y/y^4$. For that purpose we consider the simplest solution~\eqref{of9}: $b=2M=r_0$ ($m=1/2$ and $\bt=\infty$), then we asymptotically solve~\eqref{3f4} for $A$ to find
\begin{align}
&A=1-\frac{1}{y}+\frac{4 (q_1^2+q_2^2) \ln  y}{y^2}+\frac{c}{y^2}-\frac{4 (q_1^2+q_2^2) \ln  y}{y^3}\nn\\
&-\frac{c+4 q_2^2}{y^3}+\frac{s_2 \ln ^2y+s_1\ln  y+s_0}{y^4},\\
&s_0=c^2+2 c q_2^2+2 q_2^2 (1+2 q_1^2+2 q_2^2),\nn\\
&s_1=8 (c+q_2^2) (q_1^2+q_2^2),\;\;s_2=16 (q_1^2+q_2^2)^2.\nn
\end{align}
Here $c$ is a function of $q_2^2$ such that $\lim_{q_2^2\to 0}c=1$. The first four terms (up to $1/y^2$) are enough to yield
\begin{equation}\label{de1}
r_0^2\om=a\Big(\frac{1}{y^3}-\frac{4(q_1^2+q_2^2)\ln y}{y^4}-\frac{c}{y^4}+\cdots\Big),
\end{equation}
as in~\eqref{tf18}. Thus, the charges contribute additively to the dragging effects: the more charges one adds to the solution the lower the dragging effects on falling neutral objects become. The Kerr-Newman black hole is not endowed with such a property: the contribution of the charges is also additive but the $\ln r$ factor, diverging at spatial infinity, is missing. Its angular velocity is given by
\begin{equation*}
\om_{\text{K-N}}=a\Big(\frac{2M}{r^3}-\frac{Q^2}{r^4}+\cdots\Big),
\end{equation*}
In~\eqref{de1}, while the charges are bounded from above, their number may be augmented at will. Moreover, their contribution is proportional to $\ln y/y^4$. Observation of falling neutral objects may constitute a good substitute for known techniques used for distinguishing rotating black holes from rotating wormholes that are harbored in the center of galaxies.

Instead of~\eqref{3f2} and~\eqref{3f3} we could make the following choice:
\begin{align*}
&\bar{\ep}=-\bar{p}_r=\bar{p}_{\ta}=\bar{p}_{\phi}\equiv \frac{Q_1^2}{8\pi\ro^4},\\
&\tilde{\ep}=-\tilde{p}_r=\tilde{p}_{\ta}\equiv \frac{Q_2^2}{8\pi\ro^4},
\end{align*}
then determine $\tilde{p}_{\phi}$ upon solving S2. This would lead to the same constraint~\eqref{3f4} and the same metric solution~\eqref{rmq1q2} but the expression of $\tilde{p}_{\phi}$ would be different from the rhs of~\eqref{3f7}.

We could also make the following choice:
\begin{align*}
&-\bar{p}_r=\bar{p}_{\ta}=\bar{p}_{\phi}\equiv \frac{Q_1^2}{8\pi\ro^4},\\
&\tilde{\ep}=-\tilde{p}_r=\tilde{p}_{\ta}=\tilde{p}_{\phi}\equiv \frac{Q_2^2}{8\pi\ro^4},
\end{align*}
then solve for $\bar{\ep}$.

\section{The weak energy condition\label{secwec}}

In Ref.~\cite{Trot} we have shown that the null energy condition and the weak energy condition (WEC) are always satisfied on paths, through the throat, located on cones of equations $\ta=\text{constant}$. We have also shown that along these conical paths, the WEC remains satisfied whether the motion on the paths is undergone in the direction of rotation of the wormhole, in the opposite one, or in both directions in a zigzag pattern.

The aim of this section is to reach similar conclusions for any generic three-fluid-sourced rotating wormhole [more generic than~\eqref{rmq1q2}; that is, without constraining $A$ and $b$ as in~\eqref{3f4}, which is the same as~\eqref{3f6}]. There are, however, two main differences: the problem at hands is more involved than the one treated in Ref.~\cite{Trot} because of (1) the presence of three fluids and (2) the non-reversibility of~\eqref{3f6}. It is not possible to first choose an expression for $b$ [preferably of the form~\eqref{of9}], as we did in Ref.~\cite{Trot}, such that its derivative $b'$ and the total energy density~\eqref{g3b} of the nonrotating counterpart wormhole are positive for all $r\geq r_0$, then determine $A$. In this work, the non-reversibility of~\eqref{3f6} forces us to first choose an expression for $A$ then determine that of $b$. But this does not always yield a function $b$ with the desired properties, as this is the case with the solution~\eqref{rmq1q2m} where $b'$ and the energy density~\eqref{g3b} of the nonrotating  wormhole have both signs. We may expect to encounter violations of the WEC, in the geometry of the rotating wormhole~\eqref{rmq1q2m}, even on the above-mentioned conical paths as their nonrotating counterparts do violate the WEC.

In the most generic case of a three-fluid-sourced rotating wormhole, the physical WEC is the constraint
\begin{equation}\label{wec1}
    W\equiv (T_{\mu\nu}+\bar{T}_{\mu\nu}+\tilde{T}_{\mu\nu})u^{\mu}u^{\nu}\geq 0,
\end{equation}
expressing the positiveness of the local energy as seen by any timelike vector $u^{\mu}$ ($u^{\mu}u_{\mu}=1$). Using the basis~\eqref{of1}, this is of the form~\eqref{of3a}
\begin{align}\label{wec2}
&u^{\mu}=U(e^{\mu}_t+s_1e^{\mu}_r+s_2e^{\mu}_{\ta}+s_3e^{\mu}_{\phi}),\nn\\
&U=\frac{1}{\sqrt{1-s_1^2-s_2^2-s_3^2}},\\
&-1<s_i<1\quad\text{ and }\quad\sum_{i=1}^3s_i^2<1\qquad (i:1\to 3).\nn
\end{align}

Recall that in the nonrotating case ($P_{\ta}=P_{\phi}=P_t$), the WEC is expressed as
\begin{equation}\label{wec3}
    W_{\text{NR}}=E+S_1^2P_r+S_2^2P_t\geq 0
\end{equation}
($S_1^2=s_1^2<1$ and $S_2^2=s_2^2+s_3^2<1$), where $P_r=p_r+\bar{p}_r+\tilde{p}_r$ and $P_t=p_{\theta }+\bar{p}_{\theta }+\tilde{p}_{\theta }$ are the total radial and transverse pressures and $E=\ep+\bar{\ep}+\tilde{\ep}$ is the total energy density. Since $S_1$ and $S_2$ are arbitrary, this results in
\begin{equation}\label{wec4}
   E\geq0,\quad E+P_r\geq0,\quad E+P_t\geq0.
\end{equation}

In the rotating case, upon using~\eqref{of4a}, \eqref{tf3}, \eqref{3f1} and~\eqref{wec2} we express the WEC~\eqref{wec1} in its general form as
\begin{align}\label{wec5}
&W=W_t+s_1^2W_r+s_2^2W_{\ta}+s_3^2W_{\phi}+s_3W_{t\phi}\geq 0,\nn\\
&W_t=\epsilon +\frac{(a^2+r^2)^2 \bar{\epsilon}+a^2 \Delta  \bar{p}_{\phi }\sin ^2\theta }{\Si }
+\frac{\Delta  \tilde{\epsilon }+a^2 \tilde{p}_{\phi } \sin ^2\theta }{\Delta _{\theta}},\nn\\
&W_r=p_r+\bar{p}_r+\tilde{p}_r,\qquad W_{\theta }=p_{\theta }+\bar{p}_{\theta }+\tilde{p}_{\theta },\\
&W_{\phi }=p_{\phi }+\frac{a^2 \Delta  \bar{\epsilon} \sin ^2\theta +(a^2+r^2)^2 \bar{p}_{\phi }}{\Si }
+\frac{a^2 \tilde{\epsilon } \sin ^2\theta +\Delta  \tilde{p}_{\phi }}{\Delta _{\theta }},\nn\\
&W_{t\phi}=2 a \sqrt{\Delta } \Big(\frac{(a^2+r^2) (\bar{\epsilon}+\bar{p}_{\phi })}{\Si }+\frac{\tilde{\epsilon }+\tilde{p}_{\phi }}{\Delta _{\theta }}\Big) \sin  \theta ,\nn
\end{align}
regardless of the particular three-fluid rotating wormhole. This applies too to the one- and two-fluid-sourced rotating wormholes derived in this work. Now, since ($s_1,s_2,s_3$), as defined in~\eqref{wec2}, are arbitrary, this results in
\begin{multline}\label{wec6}
   W_t\geq0,\quad W_t+W_r\geq0,\quad W_t+W_{\theta }\geq0,\\
   W_t+W_{\phi}+W_{t\phi}\geq0,\quad W_t+W_{\phi}-W_{t\phi}\geq0.
\end{multline}
Both~\eqref{wec5} and~\eqref{wec6} reduce to~\eqref{wec3} and~\eqref{wec4} if rotation is suppressed ($a\equiv0$). The signs of the $W$'s are not constant on the whole range of ($r,\ta,a^2$). Depending on the sign of $W_{t\phi}$, the last two conditions~\eqref{wec6} imply each other in the one or the other way.

We see that each $W$~\eqref{wec5} is the sum of three terms, the first of which is due to exotic matter ($T_{\mu\nu}$) and the two others are due to ordinary matters ($\bar{T}_{\mu\nu}$, $\tilde{T}_{\mu\nu}$). The second and third terms in each expression $W$ can be made positive by judicious choices of the SETs $\bar{T}_{\mu\nu}$ and $\tilde{T}_{\mu\nu}$, as we did in the previous section where ($\bar{\epsilon},\bar{p}_{\phi }$) and ($\tilde{\epsilon},\tilde{p}_{\phi }$) were taken positive. The contribution of these two SETs, if judiciously chosen, is to confine the effects of the exotic matter, generated by $T_{\mu\nu}$, and alleviate the violation of the WEC.

The expressions of $W_t$ and $W_{\phi}$ may be arranged as
\begin{align}\label{wec7}
&W_t=\epsilon +\bar{\epsilon}+\tilde{\epsilon}+\frac{a^2\Delta (\bar{\epsilon}+\bar{p}_{\phi })\sin ^2\theta }{\Si }
+\frac{a^2 (\tilde{\epsilon}+\tilde{p}_{\phi })\sin ^2\theta }{\Delta _{\theta}},\nn\\
&W_{\phi }=p_{\phi }+\bar{p}_{\phi }+\tilde{p}_{\phi }+\text{same terms}.
\end{align}
Here the sums $\epsilon +\bar{\epsilon}+\tilde{\epsilon}$ and $p_{\phi }+\bar{p}_{\phi }+\tilde{p}_{\phi }$ are not the purely nonrotating contributions, for the components of the three SETs depend on $a^2$. However, in the limit of slow rotation, these sums approach their nonrotating values and the additional terms in~\eqref{wec7}, proportional to $\sin^2\ta$, serve to alleviate the violation of the WEC of the nonrotating case if $\bar{\epsilon}+\bar{p}_{\phi }$ and $\tilde{\epsilon}+\tilde{p}_{\phi }$ are positive. These constraints are weaker than those discussed in the previous paragraph.

As we mentioned earlier, for a generic three-fluid-sourced rotating wormhole we expect to see the WEC violated, so we will not examine the conditions of its fulfilment~\eqref{wec6}; rather, we will seek conical paths ($s_2\equiv 0$) through the throat along which the WEC~\eqref{wec5} is satisfied. To be more specific, we will determine the \emph{necessary} conditions for such paths to exist; that is, we mostly focus on the region near the throat. The determination of the necessary and sufficient conditions is analytically involved problem and could only be solved numerically.

We too restrict ourselves to the slow rotation limit
\begin{equation}\label{wec8}
    r_0\om(r_0)=\frac{2ar_0f(r_0)}{\Si(r_0)}=\frac{ar_0^3[1-A(r_0)]}{\Si(r_0)}\ll 1,
\end{equation}
which states that the linear velocities of dragged objects approaching the throat are much smaller than the speed of light. This ensures safe traversability. This limit implies $r\om(r)\ll 1$ since $r\om(r)$ is a decreasing function of $r$. Setting $s_2=0$, the condition~\eqref{wec5} reads in the slow rotation limit
\begin{equation}\label{wec9}
W/U^2=\al s_3^2+a\bt s_3+\ga\geq0,
\end{equation}
where ($\al,\bt,\ga$) do not depend on $a$ and $\ga$ depends on $s_1^2$. When the roots of $\al s_3^2+a\bt s_3+\ga =0$ are real, they are given by
\begin{equation*}
s_{3\pm }=\pm \sqrt{-\frac{\ga}{\al}}-\frac{a\bt}{2\al}+\mathcal{O}(a^2).
\end{equation*}

Table~\ref{Tab1} shows the generic cases where the conical paths $s_2\equiv 0$, along which the WEC is satisfied, exist. Some of these cases may not be realizable, depending on the specific three-fluid rotating wormhole. For instance, it can be shown that the case 1a ($\al<0$ and $-\ga/\al<0$) is not realizable if the solution is the three-fluid-sourced rotating wormhole~\eqref{rmq1q2m}.
\begin{table}[h]
{\footnotesize
\begin{tabular}{|c|c|c|c|c|c|}
  \hline
  &  &  &  &  &  \vspace{-0.3cm} \\
  {Case} & ${\al}$ & ${-\frac{\ga}{\al}}$ & ${-\frac{\ga}{\al}-1}$ & {WEC} &  for all \\
  &  &  &  &  &  \vspace{-0.3cm} \\
  \hline
  \hline
  1a & $-$ & $-$ & $-$ & V & $-1<s_3<1$  \\
  \hline
  1b & $-$ & $+$ & $-$ & S & $-1<s_{3-}<s_3<s_{3+}<1$  \\
  \hline
  1c & $-$ & $+$ & $+$ & S & $-1<s_3<1$  \\
  \hline
  2a & $+$ & $-$ & $-$ & S & $-1<s_3<1$  \\
  \hline
  2b & $+$ & $+$ & $-$ & S & $-1<s_3<s_{3-}$ or $s_{3+}<s_3<1$  \\
  \hline
  2c & $+$ & $+$ & $+$ & V & $-1<s_3<1$  \\
  \hline
\end{tabular}}
\caption{{\footnotesize Existence of conical paths where the physical WEC is satisfied. Note that in the cases 1b, 1c and 2a the motion on the paths may be undergone in the direction of rotation of the wormhole, in the opposite one, or in both directions in a zigzag pattern since $s_3$ may have both signs, while for the case 2b the motion is undergone in the one or the other direction. \textsc{Nomenclature:} ``S" for ``Satisfied" and ``V" for ``Violated".}}\label{Tab1}
\end{table}

The conical paths along which the WEC is satisfied may exist for different cases (cases 1b to 2b, as shown in Table~\ref{Tab1}) constraining ($\al,\ga$). When expressed in terms of the charges ($m,q_1,\dots$) and $s_1$, each case splits into sub-cases where each sub-case appears to be a set of inequalities and equalities constraining ($m,q_1,\dots,s_1$).

\section{Generating ($\pmb{n+1}$)-fluid-sourced, $\pmb{n}$-charged, rotating wormholes\label{secnf}}

In the previous section we dealt with the problem where the SET of the total matter content is the sum $T^{\mu\nu}+\bar{T}^{\mu\nu}+\tilde{T}^{\mu\nu}$ of three sub-SETs with $T^{\mu\nu}$ being that of an exotic matter and the other two correspond to electromagnetic matter contents. There are two other possibilities: we could work out the problem where $\bar{T}^{\mu\nu}$ (resp. $\tilde{T}^{\mu\nu}$) is taken as exotic. However, our experience with the two-fluid-sourced rotating wormholes, treated in Sec.~\ref{sectf}, prevents us from doing so, for these configurations might be much involved to be treated analytically.

To each frame $e$~\eqref{of1}, $\bar{e}$~\eqref{of12}, and $\tilde{e}$~\eqref{of13} we associated a sub-SET. Continuing this way we may be able to construct ($n+1$)-fluid-sourced, $n$-charged, rotating wormholes by choosing $n+1$ frames.

The frames $e$, $\bar{e}$, and $\tilde{e}$ have been constructed based on the following decomposition of the $t\phi$ part of the metric:
\begin{align}\label{nf1}
\dd s_{t\phi}^2 =& \Big(1-\frac{2f}{\ro^2}\Big)\dd t^2+\frac{4af \sin ^2\theta}{\ro^2}\,\dd t\dd \phi-\frac{\Si\sin ^2\theta}{\ro^2}\,\dd \phi^2\nn\\
=&(f_1\dd t+f_2\dd \phi)^2-(f_3\dd t+f_4\dd \phi)^2,
\end{align}
where ($f_1>0,f_2,f_3,f_4>0$) are functions of ($r,\ta$), which provide the corresponding 1-forms. Inspection of~\eqref{rm1}, \eqref{rm2}, and~\eqref{rm3} reveals the expressions of ($f_1,f_2,f_3,f_4$) for the frames $e$, $\bar{e}$, and $\tilde{e}$ respectively.

To construct new frames one needs to find new sets ($f_1>0,f_2,f_3,f_4>0$), which are solutions to
\begin{align}
&f_1^2-f_3^2=\Big(1-\frac{2f}{\ro^2}\Big),\nn\\
\label{nf2}&f_1f_2-f_3f_4=\frac{2af \sin ^2\theta}{\ro^2},\\
&f_4^2-f_2^2=\frac{\Si\sin ^2\theta}{\ro^2}.\nn
\end{align}
One can fix an expression, say, for $f_i$ then determine the rest of the functions $f_j$ ($j\neq i$) upon solving~\eqref{nf2}. One can also fix an expression for some ratio $f_i/f_j$ that brings the number of equations equal to that of the unknown functions ($f_1,f_2,f_3,f_4$). For those functions ($f_1>0,f_2,f_3,f_4>0$) that remain defined on the whole range of ($r,\ta$), a new frame $\hat{e}$ is associated to the set of one 1-forms: $\hat{\om}^t\equiv (f_1\dd t+f_2\dd \phi)$, $\hat{\om}^r\equiv \om^r$, $\hat{\om}^{\ta}\equiv \om^{\ta}$, $\hat{\om}^{\phi}\equiv -(f_3\dd t+f_4\dd \phi)$. The 1-forms ($\om^r,\om^{\ta}$) are defined in the first paragraph of Sec.~\ref{secof}.

The total matter content is now the sum $T^{\mu\nu}+\bar{T}^{\mu\nu}+\tilde{T}^{\mu\nu}+\hat{T}^{\mu\nu}$ with $\hat{T}^{\mu\nu}=\hat{\ep} \hat{e}^{\mu}_t\hat{e}^{\nu}_t+\hat{p}_r\hat{e}^{\mu}_r\hat{e}^{\nu}_r+\hat{p}_{\ta}\hat{e}^{\mu}_{\ta}\hat{e}^{\nu}_{\ta}
+\hat{p}_{\phi}\hat{e}^{\mu}_{\phi}\hat{e}^{\nu}_{\phi}$. The resolution of the sets of field equations S1 and S2 proceeds as before: $T^{\mu\nu}$ is exotic and ($\bar{T}^{\mu\nu},\tilde{T}^{\mu\nu},\hat{T}^{\mu\nu}$) are electromagnetic with, say,
\begin{align}
&\label{nf3}\bar{\ep}=-\bar{p}_r=\bar{p}_{\ta}\equiv \frac{Q_1^2}{8\pi\ro^4},\\
&\label{nf4}\tilde{\ep}=-\tilde{p}_r=\tilde{p}_{\ta}=\tilde{p}_{\phi}\equiv \frac{Q_2^2}{8\pi\ro^4},\\
&\label{nf5}\hat{\ep}=-\hat{p}_r=\hat{p}_{\ta}=\hat{p}_{\phi}\equiv \frac{Q_3^2}{8\pi\ro^4}.
\end{align}
The values of ($p_r,p_{\ta}$) are determined upon solving the set S1 and those of ($\ep,p_{\phi},\bar{p}_{\phi}$) are determined upon solving the set S2. In order that $\lim_{a^2\to 0}\bar{p}_{\phi}$ reduces to the static value $Q_1^2/(8\pi r^4)$, we may need to constrain $A$ and $b$ as we did in~\eqref{3f4}, and so on.

\section{Conclusion \label{secc}}

We have shown in Ref.~\cite{Trot} and in this work that two types of massive, charged, rotating wormholes can be derived from the general metric~\eqref{g7}. For the wormholes derived in Ref.~\cite{Trot} it was shown that there exists a mass-charge constraint yielding almost no more dragging effects than ordinary stars. The dragging effects of the wormholes derived in this work, which by no means can mimic those of ordinary stars, differ appreciably from those of the the Kerr-Newman black hole by the presence of a logarithmic term that diverges at spatial infinity. These effects could be used as potential mean in astrophysical observations meant to investigate the nature of the supermassive black hole candidates that some galactic centers, as the Sgr A$^\star$, harbor.

The three frames used in this work are the most common ones. To each frame one can attach a form of matter. Given $n+1$ frames one can in principal construct rotating and nonrotating wormholes their SET is the sum of $n+1$ sub-SETs, $n$ of which are electromagnetic and the left one is exotic.

The static wormholes obtained in this work were not derived by direct integration; rather, they were derived as the limit $a\to 0$ of their rotating counterparts. They can hardly be derived analytically. This shows that the method introduced here, which consists in selecting different moving (here rotating) frames and attach to each frame a form of matter, a SET $T^{\mu\nu}$, constitutes a new approach of integration for both rotating and static solutions.




\end{document}